\documentclass[preprint]{aastex}

\renewcommand{\vec}[1]{\mathbf{#1}}

\begin{document}

\title{The Self-Similarity of Shear-Dominated Viscous Stirring}
\author{Benjamin F. Collins, Hilke E. Schlichting, and Re'em Sari}
\affil{California Institute of Technology, MC 130-33, Pasadena, CA 91125}
\email{bfc@tapir.caltech.edu}

\begin{abstract}

We examine the growth of eccentricities of a population of
particles with initially circular orbits around a central massive body.
Successive encounters between pairs of particles increase 
the eccentricities in the disk on average.  As long as the 
epicyclic motions of the particles are small compared to 
the shearing motion between Keplerian orbits, 
there is no preferred scale for the eccentricities. 
The simplification due to this self-similarity allows us to find 
an analytic form for the distribution function;
full numerical integrations of a disk with 200 planetesimals verify our
analytical self-similar distribution.  The shape of this non-equilibrium
profile is identical to the equilibrium profile of a shear-dominated
population whose mutual excitations are balanced by dynamical friction 
or Epstein gas drag.

\end{abstract}


\section{INTRODUCTION}

Modern computational power allows the simultaneous 
integration of the orbits of increasingly numerous particles.
Much of the planet formation process, however, 
involves particle numbers that exceed the 
limits of computational efficiency.  This limitation is often
circumvented with a statistical approach.
By monitoring the gravitational interactions of the particles
in a time-averaged sense, various properties of the particle population
can be calculated without a full $N$-body simulation.

\citet{CS06}, hereafter Paper I, motivate a Boltzmann equation to describe the 
evolution of the eccentricity distribution of an ensemble of particles in which
the relative motion between any two interacting particles is
dominated by the shearing motion of close circular orbits.  Such a regime of  
orbital eccentricities is called shear-dominated.  The solution of their equation
provides a simple analytic expression for the equilibrium eccentricity distribution 
that results when dynamical friction can balance the mutual interactions
of the particles; the analytic expression matches results from 
numerical simulations remarkably well.  

In this letter we derive analytically the non-equilibrium distribution function 
of interacting shear-dominated particles in the absence of dynamical friction.
\S \ref{secReview} reviews the construction of the Boltzmann 
equation.
In \S \ref{secAnalytical}, 
we show that the distribution function behaves self-similarly, and 
the shape of the non-equilibrium distribution function is 
identical to the equilibrium distribution of Paper I.
\S \ref{secNumerical} corroborates this result with numerical 
simulations.  Conclusions follow in \S \ref{secConclusions}.

\section{THE TIME-DEPENDENT BOLTZMANN EQUATION}
\label{secReview}

We consider a disk of particles on initially circular 
orbits around a massive central body.  We write their surface 
mass density $\sigma$ and the mass of a single body $m$.  The
number density, $\sigma/m$ is sufficiently low that three-body
encounters are very rare, therefore the orbital evolution of each 
body is well described as a sequence of pair-wise encounters.

The change in eccentricity due to one such encounter can be calculated 
analytically. 
For completeness, we summarize the derivation presented in Paper I.
Let one particle, with 
a semi-major axis $a$, encounter another with semi-major axis $a+b$.  
In the limit of $b \ll a$, 
the relative orbital frequency between the pair is  
$\Omega_r = (3/2)\Omega b/a$, where $\Omega$ is the Keplerian orbital
frequency for a semi-major axis $a$.  If in addition 
$b \gg R_H \approx (m/M_{\odot})^{1/3} a$,
 the change in eccentricity from one encounter is 
$e_k = A_k (m/M_{\odot}) (b/a)^{-2}$, 
where $A_k \approx 6.67$ collects the order-unity coefficients 
\citep{GT78,PH86}. 

The eccentricity is not the only Keplerian element that characterizes
the non-circular motion of a particle; the longitude of periapse
specifies the relative orientation of a particle's epicycle. 
The particles may also follow orbits that do not lie in the disk.
However, shear-dominated viscous stirring excites 
inclinations at a rate that is always slower than the excitation of eccentricities
\citep{WS93,GLS04,Raf03c}.  The perpendicular velocities are, in this case, 
always negligible compared to the epicyclic motion in the disk plane.  

The magnitude of an orbit's eccentricity and the longitude of periapse
together specify a two-dimensional parameter space.  We describe the
two-dimensional variable with a vector, $\vec e = \{e \cos{\omega}, e
\sin{\omega}\}$.  The distribution function is a function of this
vector and time, $f({\vec e},t)$.  That the changes in $\vec e$ due to
encounters do not depend on the longitude of periapse already shows
that the distribution function must be axisymmetric, or $f(\vec
e,t)=f(e,t)$.  Then the number of bodies per unit logarithmic interval
around $e$ is given by $2\pi e^{2}f(e,t)$.

We characterize the eccentricity growth with a differential rate,
$p(\vec e_k)d^{2}\vec e_k$, that the eccentricity vector of a particle
will be changed by an amount $\vec e_k$.  Since the change in
eccentricity experienced by a pair of bodies, when treated as a vector
quantity, is independent of the initial eccentricity vector of each
body, this function is also axisymmetric and only depends on the
magnitude of the change of eccentricity, $e_k$.

The excitation rate depends on the surface mass density of particles
in the disk, $\sigma$, the mass of a single body, $m$, the mass of the
central star, $M_{\odot}$, the cross-section at which a particle
experiences encounters of a strength $e_k$, and the relative speed of
those encounters.  The impact parameter at which a particle receives
an eccentricity $e_k$ scales as $b\sim e_k^{-1/2}$.  If the
eccentricities are small, the speed at which one particle encounters
the others is set only by the shearing of their two orbits, which is
proportional to $b$.  Then, as shown in Paper I,

\begin{equation}
2 \pi p(e_k) e_k d e_k = 3 \frac{\sigma}{m} \Omega b(e_k) db(e_k).
\end{equation}

\noindent
After simplification, we find

\begin{equation}
p(e_k) = \frac{A_k}{4\pi} \frac{\sigma a^2}{M_\odot} \frac{1}{e_k^3} \Omega.
\end{equation}

\noindent
An integral over every $e_k$ dictates the rate of change of 
the number of bodies with a given eccentricity, $e$:

\begin{equation}
\label{eqFDiffEq}
\frac{\partial f(e,t)}{\partial t} = \int\int p(|\vec e-\vec e_n|)
\left[f(e_n,t)-f(e,t)\right] d^2 \vec{e_n}.
\end{equation}
 
\noindent
Note that this equation implicitly conserves the total particle number,
$\int\int f(e,t) d^2\vec e = 1$.  This can be shown by integrating both sides
with respect to $\vec e$.

\section{THE SELF-SIMILAR DISTRIBUTION}
\label{secAnalytical}

Without a specific eccentricity scale to dictate the evolution of 
$f(e,t)$, we expect a solution of the form,

\begin{equation}
\label{eqSelfSimilarF}
f(e,t)=F(t)g\left(e/e_{c}(t)\right).
\end{equation}

\noindent
Replacing $f(e,t)$ in equation \ref{eqFDiffEq} with equation
\ref{eqSelfSimilarF}, we find,

\begin{equation}
\label{eqFandGDiffEq}
\frac{1}{F(t)}\frac{\partial F(t)}{\partial t}e_c(t) g(x)
- x \frac{\partial g(x)}{\partial x} \dot e_c(t)
=\int\int p(|\vec x-\vec x_n|)\left[g(x_n)-g(x)\right]d^2\vec {x_n},
\end{equation}

\noindent 
where $\vec x = \vec e/e_c(t)$.  The additional constraint that 
equation \ref{eqFDiffEq} conserves particle number implies 
$F(t)e_c(t)^2$ is constant.  This relationship simplifies 
the left side of equation \ref{eqFandGDiffEq} such that the 
only possible time-dependence of each term is contained in $\dot e_c(t)$.  
The right-hand side, however, is independent of time.  Therefore
$\dot e_c(t)$ must be constant. Then,

\begin{equation}
\label{eqEStarofT}
e_c(t)=C_e t ~  {\rm and}~ F(t)=(C_e t)^{-2}.
\end{equation}

\noindent
The overall normalization of $F(t)$ is arbitrary, as it can be
absorbed into $g(x)$.  Our choice of $F(t)$ requires $\int\int
g(x)d^2\vec x = 1$ to ensure that $\int\int f(e,t)d^2\vec e =1$ for
all $t$.  Physically, the typical eccentricity, $e_c(t)$, is set by
the eccentricity change that occurs once per particle per time $t$, or,
$e_c(t)^2p(e_c(t))t \sim 1$.  This argument sets $e_c(t)$ only up to a
constant coefficient; for simplicity we choose the coefficients such
that $e_c(t) = (A_k/2)(\sigma a^2 /M_{\odot}) \Omega t$.  Then, the
profile shape, $g(x)$, is specified by the integro-differential
equation

\begin{equation}
\label{eqGofXDiffEq}
2 g(x)
+x \frac{\partial g(x)}{\partial x} 
+ \frac{1}{2 \pi} \int\int \frac{g(x_n)-g(x)}{|\vec x_n - \vec x|^3} d^2\vec {x}_n = 0,
\end{equation}

\noindent
Equation \ref{eqGofXDiffEq} is identical to equation 17 of Paper I.
A detailed description of the equation and a proof of its solution 
can be found in that paper.  For reference, the solution is

\begin{equation}
\label{eqGofX}
g(x) = \frac{1}{2 \pi} \left(1+x^2\right)^{-3/2},
\end{equation}

\section{NUMERICAL SIMULATIONS}
\label{secNumerical}

The non-equilibrium distribution function of eccentricities in the
regime discussed above can be measured directly from a full numerical
simulation of the disk.  We use a custom N-body integrator that
evolves the changes in the two-body constants of motion of each
particle around the central mass.  These constants are chosen to vary
slowly with small perturbations.  Solving Kepler's equation for each
body translates each time-step into a change in orbital phase.  The
constants of motion are then integrated by a fourth-order Runge-Kutta
routine with adaptive time-steps \citep{numericalrecipies}.

For this study we follow a disk of two-hundred equal mass bodies, with
$m=5\times 10^{-9} M_{\odot}$, on initially circular orbits with
randomly determined phases and semi-major axes within a small annulus
of width $\Delta a = 0.8 a$.  To avoid possible artifacts from the
edge of the simulation, we only measure the eccentricities of the
bodies in the central third of the disk.  A histogram of those
eccentricities shows the number of bodies
with each eccentricity, $e~ dN/de$.  To increase the signal to noise
ratio of the histogram at each time, we add the results of one hundred
simulations with randomly generated initial semi-major axes and
orbital phases.

Figure 1 shows the eccentricity distributions measured after three and
ten orbits.  The horizontal error bars indicate the width of each bin,
and the vertical error bars are determined assuming that each bin is
Poisson distributed.  The analytic distribution function derived in \S
\ref{secAnalytical} for each time is also plotted, as a solid line.
The measured distributions agree remarkably with the analytic result.

\begin{figure}
\label{figManyTimes}
\centering{\includegraphics[angle=-90,width=0.9\columnwidth]{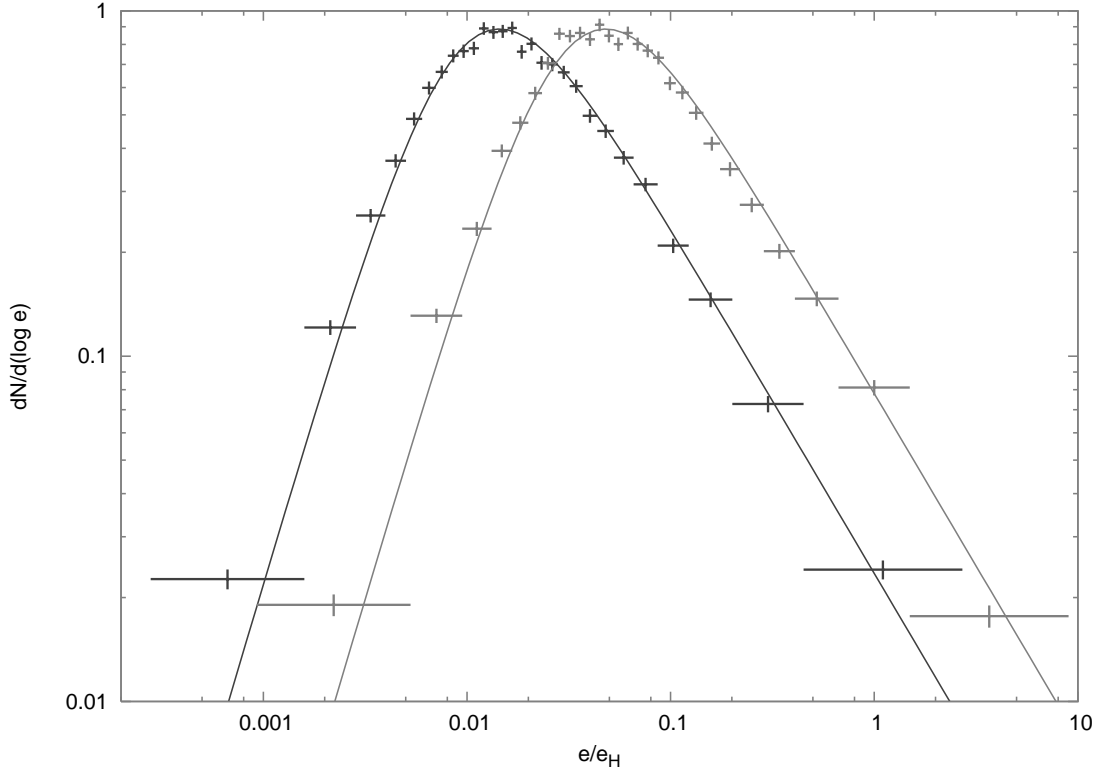}}

\caption{Eccentricity distributions of a shear-dominated disk 
of 200 particles each with a mass $m=5\times 10^{-9} M_{\odot}$
after three (black line) and ten (gray line) orbits.
The average surface mass density of the simulated annulus is
$3\times 10^{-3} {\rm g~cm^{-2}}$.  
The vertical error bars are estimated by assuming each bin obeys
Poisson statistics.
The width of each bin has been 
chosen such that each bin contains a similar number of particles.}

\end{figure}

To emphasize the self-similarity of the distribution shape, we scale
the eccentricities measured at each time by the characteristic eccentricity
at that time ($e_c(t)$, given by equation \ref{eqEStarofT}) and plot the 
shapes added together.  Figure 2 shows that the resulting
distribution shape matches the analytic form of $g(x)$ very well.

\begin{figure}
\label{figScaledDist}
\centering{\includegraphics[angle=-90,width=0.9\columnwidth]{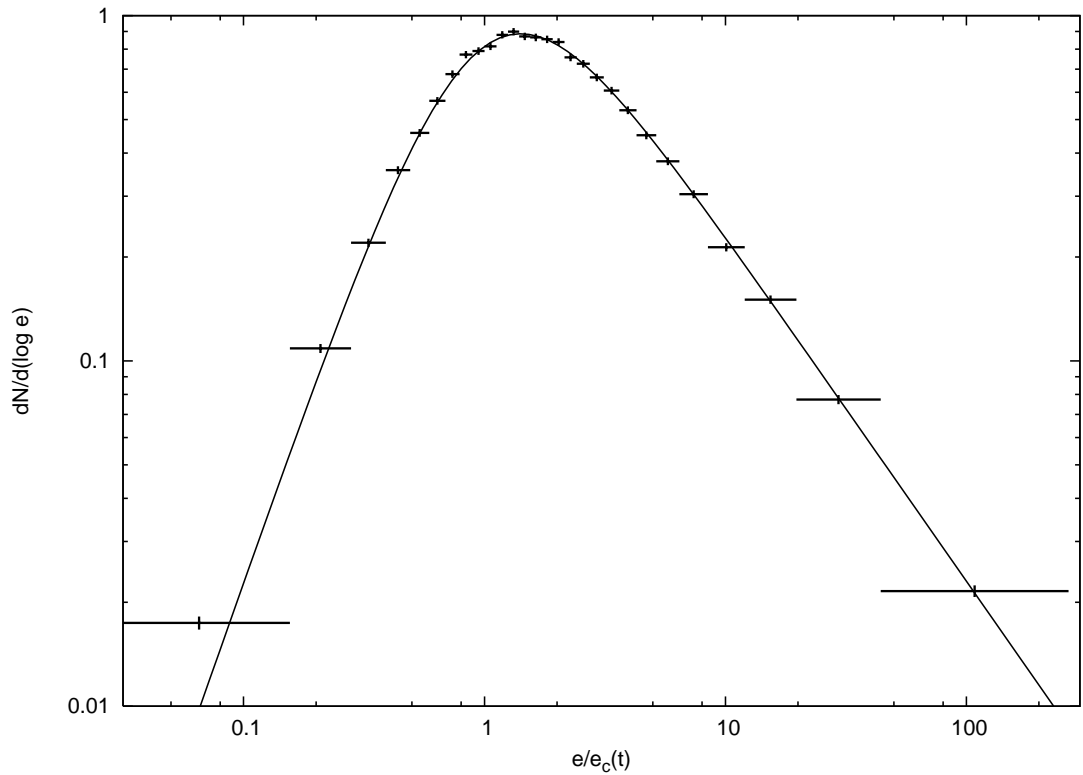}}

\caption{The eccentricity distribution of the same 
numerical simulation of Figure 1.  Here, the distribution after
one, three, and ten orbits, scaled by the characteristic eccentricity 
at that time are added together.  The
profile shape is very well described by equation \ref{eqGofX},
plotted as a solid line.  The error bars are chosen in the same way as Figure 1.}

\end{figure}

\section{DISCUSSION}
\label{secConclusions}

We have written a time-dependent Boltzmann equation that 
describes the eccentricity distribution function of a 
population of orbiting particles under the influence of their
mutual excitations in the shear-dominated regime.  
Reasoning that the distribution function of eccentricities 
should behave self-similarly, that is, retain a constant profile
while its normalization and scaling depend on time, we have reduced 
the Boltzmann equation to a form that has already been shown to 
have an analytic solution.  Numerical experiments confirm the self-similarity
and the analytic solution.

Although we have only considered disks of a single particle size,
the formalism above applies trivially to disks with mass distributions.
In fact, the characteristic eccentricity, equation \ref{eqEStarofT}, 
depends only on the total surface mass density of the disk.  
As long as bodies of every part of the mass spectrum are in the 
shear-dominated regime, the eccentricities of all bodies are 
drawn from the same distribution.  This is a consequence of the fact
that gravitational acceleration is mass-independent.  In contrast, 
the dynamical friction of Paper I depends on the size of each 
particle.  The equilibrium distributions in that case do differ for each
mass group.

Since $e_c(t)$ is an increasing function of time,
the condition of shear-dominated dynamics will be violated eventually.  
While the disk is shear-dominated however, most of the disk bodies
have eccentricities of about $e_c(t)$.  The mean 
eccentricity, $\int\int e f(e,t) d^2 \vec e$, is formally infinite; in reality
the mean depends logarithmically on the maximum eccentricity 
achievable from one interaction.  Higher moments of the distribution,
such as $\langle e^2 \rangle$, are dominated by the bodies with
the maximum eccentricity.  The random kinetic energy of the disk bodies,
for example, is then set by the few bodies with the highest eccentricities
regardless of the value of $e_c(t)$.

That the shape of the distribution function is identical to 
the distribution function of Paper I is ultimately not 
surprising.  In both scenarios the bodies in question excite 
their orbital parameters via the same shear-dominated viscous stirring
mechanism.
If dynamical friction is acting on these bodies, their eccentricities
decrease with time proportionally to their magnitude.  An equilibrium
between excitations and this damping produces a characteristic eccentricity
around which the eccentricities of all bodies are distributed. Without 
an agent of dynamical friction, the typical eccentricity of a body in the disk,
$e_{\rm typical} \sim e_c(t)$, grows with time. 
However, the ratio of the eccentricity of a particle that has not interacted recently, 
$e$, to that typical eccentricity shrinks proportionally to itself:

\begin{equation}
\dot x \equiv \frac{d}{dt} \left( \frac{e}{e_c(t)} \right) 
\sim - x .
\end{equation}

\noindent
This is formally equivalent to the damping provided by the 
dynamical friction of Paper I.

\bibliographystyle{apj}
\bibliography{ms}

\end{document}